\documentclass[aps,pre,preprint,superscriptaddress,showpacs]{revtex4-1}

\usepackage{graphicx}

\begin{document}


\title{Analytical Solution of Metapopulation Dynamics in Stochastic Environment}


\author{Satoru Morita}
\email[]{morita@sys.eng.shizuoka.ac.jp}
\affiliation{Department of Systems Engineering, Shizuoka University, Hamamatsu, 432-8561, Japan}
\author{Jin Yoshimura}
\email[]{jin@sys.eng.shizuoka.ac.jp}
\affiliation{Department of Systems Engineering, Shizuoka University, Hamamatsu, 432-8561, Japan}
\affiliation{Marine Biosystems Research Center, Chiba University, Uchiura, Amatsu-Kominato, Chiba 299-5502, Japan}
\affiliation{Department of Environmental and Forest Biology, State University of New York College of Environmental Science and Forestry, Syracuse, NY13210, USA}


\date{\today}

\begin{abstract}
We study a discrete stochastic linear metapopulation model to understand 
the effect of risk spreading by dispersion. We calculate analytically the 
stable distribution of populations in different habitats. The simultaneous 
distribution of populations in habitats has a complicated self-similar 
structure, but the population in each habitat follows a log-normal 
distribution. A class of discrete stochastic matrix models were 
mostly dealt numerically. 
Our analytical predictions are robust in the wide range of parameters.
Qualitative predictions of the current results should 
hold in the case of multiple habitats.
We thus conclude that environmental stochasticity always promotes dispersal.
\end{abstract}

\pacs{87.10.Mn, 87.23.Cc, 89.65.Gh}

\maketitle


Environmental heterogeneity and flucuations are 
both important in ecological and evolutionary biology \cite{r1,r2}. 
Assuming that populations face time-varying environmental 
conditions, environmental heterogeneity often gives rise to paradoxical
behavior \cite{r3}. 
For example, Jensen and Yoshimura \cite{r4} present a discrete-time 
model of offspring allocation into two habitats, either of which is 
so poor that populations cannot survive if the two habitats are
exclusive. 
The authors demonstrate that dispersal can lead to population 
persistence.
Similar paradoxical behaviors have been attracting attention in various 
research fields such as ecological biology \cite{r5,r6,r7,r8}, 
financial economics \cite{r9,r10,r11} and information engineering \cite{r12}.
In this paper, we study a discrete stochastic linear metapolulation
model \cite{r7}, which is given as discrete-time stochastic matrix models. 
This class of models have been mostly dealt numerically by simulations, 
because of the analytical difficulty in tracking the stochastic processes. 
Here, we present an analytical method and show that these models are 
principally tractable analytically.
To characterize long-term behavior of the entire system, geometric 
mean of local growth rate has been often used \cite{r3,r4,r11}. 
However, this simple approach is not valid except for well-mixed cases.
We present an accurate method to assess the long-term growth in
this model. 
We derive analytically the stable distribution of populations.

Let us consider populations that inhabits in $n$
discrete habitats. Let $x_i(t)$ be the number of individuals in 
patch $i$ at time $t$. 
Thus, the state of the populations is described by a vector 
${\bf x}_i(t)=(x_1(t),x_2(t),\dots x_n(t))^{\mbox{\footnotesize T}}$,
where the superscript T represents the transpose.
In each habitat, the population reproduces at random growth rates. 
Then a fraction of the population disperses from a habitat to 
another habitat. 
The population dynamics is given by a discrete-time 
stochastic matrix model 
\begin{equation}
 {\bf x}(t+1)=D M(t){\bf x}(t),
\label{e1}
\end{equation}
where $D$ is a time-independent dispersal matrix, 
and $M(t)$ is a time-dependent diagonal matrix of local growth rate. 
In this paper, we focus on the case of two habitats for simplicity, 
but the results can be extended to a general case (i.e., $n>2$). 
The dispersal matrix is given as
\begin{equation}
  D=\left(
  \begin{array}{cc}
    1-q & q s\\
    q s & 1-q
  \end{array}\right).
\label{e2}
\end{equation}
Here the parameters $q$ and $s$ are between 0 and 1. 
The migration rate $q$ represents the proportion of population 
that migrates from a habitat to the other habitat. 
When $q=0$, the two habitats are isolated completely. 
The parameter $s$ is survival rate during the transportation 
between habitats. 
The matrix of growth rate is written as
\begin{equation}
  M(t)=\left(
  \begin{array}{cc}
    m_1(t) & 0\\
    0 & m_2(t)
  \end{array}\right).
\label{e3}
\end{equation}

We assume that the local growth rate $m_i(t)$ ($i=1 \mbox{ or } 2$) is 
a stochastic variable which takes one of two values, $m_-$ with probability $p$
and $m_+$ with probability $1-p$. 
Here, we set $m_-<m_+$.
This stochastic fluctuation of the local growth rate comes from 
environmental fluctuations. 
Here, we neglect the temporal correlation of the local growth rates.
But we take into account the correlation between two habitats.
We denote by $c$ the correlation coefficient between  $m_1(t)$ and $m_2(t)$ at
same time.
If we denote by $p_{++}$ the probability that $m_1(t)=m_+$ and  $m_2(t)=m_+$,
 we obtain \begin{equation}
   \begin{array}{lcl}
     p_{++}&=&(1-p)^2+cp(1-p) \\
     p_{--}&=&p^2+cp(1-p) \\
     p_{+-}&=&(1-c)p(1-p)\\
     p_{-+}&=&(1-c)p(1-p).
\end{array}
\label{e5}
\end{equation}
The initial populations at $t=0$ is set as 
$(x_1(0),x_2(0))^{\mbox{\footnotesize T}}=(1,1)^{\mbox{\footnotesize T}}$.

To solve eq.~(\ref{e1}), we consider the dynamics of the ratio of populations in two habitats
\begin{equation}
  r(t)=\frac{x_2(t)}{x_1(t)}.
\label{e6}
\end{equation}
The dynamics of $r(t)$ is written as 
\begin{equation}
  r(t+1)=f\left(\frac{m_2(t)}{m_1(t)} r(t)\right),
\label{e7}
\end{equation}
where $f(r)$ is defined as
\begin{equation}
  f(r)=\frac{r(1-q)+qs}{rqs+1-q} .
\label{e8}
\end{equation}
Equation (\ref{e7}) indicates that the future value $r(t+1)$ depends
 only on the present value $r(t)$, that is, $r(t)$ follows 
one-dimensional Markov process. 
Thus, statistical state of $r(t)$ is characterized by its 
density distribution $\rho_t(r)$. 
The evolution of $\rho_t(r)$ is described by the Frobenius-Perron
equation \cite{r15,r16}. 
Using the probability $p_{--}$, $p_{+-}$, $p_{++}$ defined in eq.~(\ref{e5}),
we have
\begin{equation}
\begin{array}{lcl}
\rho_{t+1}(r)&= & \displaystyle
 p_{+-}\frac{m_+}{m_-}g(r)\rho_t\left(\frac{m_+}{m_-}f^{-1}(r)\right)\\
& & \displaystyle
+p_{+-}\frac{m_-}{m_+}g(r)\rho_t\left(\frac{m_-}{m_+}f^{-1}(r)\right)\\
& & \displaystyle
+(p_{++}+p_{--})g(r)\rho_t(f^{-1}(r)),
\end{array}
\label{e9}
\end{equation}
where $g(r)$ is the derivative function of $f^{-1}(r)$: 
\begin{equation}
  g(r)=\frac{1-2q-q^2(s^2-1)}{(rqs+q-1)^2} .
\label{e10}
\end{equation}
The density distribution $\rho_t(r)$
converges to the stationary distribution $\rho_*(r)$ (or invariant measure)
eventually, regardless of where it begins. 
The distribution $\rho_*(r)$ is determined by the recursive 
formula obtained by substituting $\rho_*(r)$ for $\rho_t(r)$ and 
$\rho_{t+1}(r)$ in eq.~(\ref{e9}). 
Figure 1(a) shows an example of $\rho_*(r)$ with a logarithmic scale. 
This distribution $\rho_*(r)$ has a complicated 
self-similar structure. 
Figure 1(b) shows the time distribution obtained by simulation 
over 100000 time steps for the same parameter of Fig.~1(a). 
These two distributions agree exactly. 
Because of the ergodicity of a Markov process, 
the time average is equal to the ensemble average. 
The distribution of $\ln r(t)$ is symmetric around 
the vertical axis $\ln r(t)=0$ (Fig.~1), 
because this model is invariant under the permutation of $r(t)$ and $1/r(t)$.
This means that the distribution of $r(t)$ coincides with that of $1/r(t)$. 
In the special case $qs=1-q$, $r(t)$ becomes one deterministically for $t>0$,
that is, $\rho_*(r)=\delta(r-1)$. 
This case represents well-mixed populations, 
which were examined by prior theoretical works \cite{r4,r7,r13,r14}.
\begin{figure}
\includegraphics[width=\linewidth]{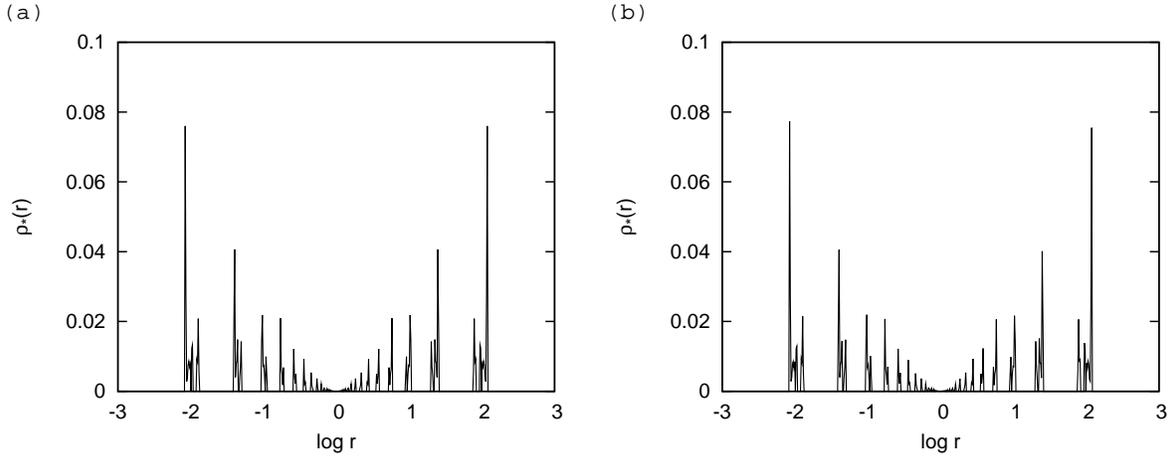}%
\caption{Stationary distribution of $r(t)=x_2(t)/x_1(t)$ for $m_+=3$,
$m_-=0.01$, $c=0$, $p=0.3$, $q=0.2$, and $s=0.5$. 
The horizontal axis is a logarithmic scale. 
(a) Theoretical result obtained by solving the Frobenius-Perron equation 
(\ref{e9}) numerically. 
(b) Time distribution obtained by computer simulation 
over 100000 time steps after a transient of 1000 time steps. 
We observe a good agreement between these two distributions. \label{f1}}
\end{figure}

The dynamics of the populations $x_1(t)$ and $x_2(t)$ 
is described as 
\begin{eqnarray}
  x_1(t+1)&=&x_1(t)[m_1(t)(1-q)+m_2(t)q s r(t)],
  \label{e11} \\
  x_2(t+1)&=&x_2(t)[m_2(t)(1-q)+m_1(t)q s /r(t)] .
  \label{e12}
\end{eqnarray}
A very important point is that the stochastic variable $r(t)$ is determined 
by eq.~(\ref{e9}) and can be treated independent of $x_1(t)$ and $x_2(t)$.
Thus, eqs.~(\ref{e11}) and (\ref{e12}) are regarded as random multiplicative
processes for $x_1(t)$ and $x_2(t)$, respectively. 
Taking into account the symmetry of $r(t)$ and $1/r(t)$,
it is obvious that the two processes (\ref{e11}) and (\ref{e12}) are identical.
Hence, it is sufficient to consider only (\ref{e11}). 
Taking the logarithms of both sides of eq.~(\ref{e11}) and 
summing them from $t=0$ to $t=T-1$, we obtain
\begin{equation}
\begin{array}{lcl}
\ln x_1(T)&= & \displaystyle
 \sum_{t=0}^{t=T-1}\ln x_1(t+1)-\ln x_1(t)\\
 &=& \displaystyle
 \sum_{t=0}^{t=T-1}\ln[m_1(t)(1-q)+m_2(t) q s r(t)] .
\end{array}
\label{e13}
\end{equation}
Thus, $\ln x_1(T)$ is given by the sum of the time series of 
the effective growth rate $\ln[m_1(t)(1-q)+m_2(t) q s r(t)]$. 
Because the effective growth rate depends on $r(t)$, 
the effective growth rates have a temporal correlation. 
However, the auto-correlation function of the effective growth rates 
decays rapidly, as shown in Fig.~2(c). 
Thus, applying the central limit theorem, we expect the distribution of
$\ln x_1(T)$ after a long time ($T\gg 1$) approaches a normal distribution.
Thus, $x_1(T)$ follows a log-normal distribution. 
The mean value of $\ln x_1(T)$ is calculated by using the stationary 
distribution of $r(t)$. 
Moreover, in an approximation in which the temporal correlation of $r(t)$ 
is neglected, we can also estimate the variance of $\ln x_1(T)$. 
In Figs.~2(a) and (b), we show examples of the evolution of the actual 
ensemble distribution and the approximation results. 
The averages (or peaks) of the actual distributions agree exactly with 
the theoretical results. 
The actual variances are also equal to the theoretical results
(Fig.~2(b)) when the auto-correlations are negligible (Fig.~2(c)). 
However, the actual variances are smaller than the approximation 
results (Fig.~2(a)), when the auto-correlations are not neglected
 (Fig.~2(c)). Here the deviations are universally smaller 
because the growth rates tend to have a negative correlation in 
short-range intervals of time. 

\begin{figure}
\includegraphics[width=\linewidth]{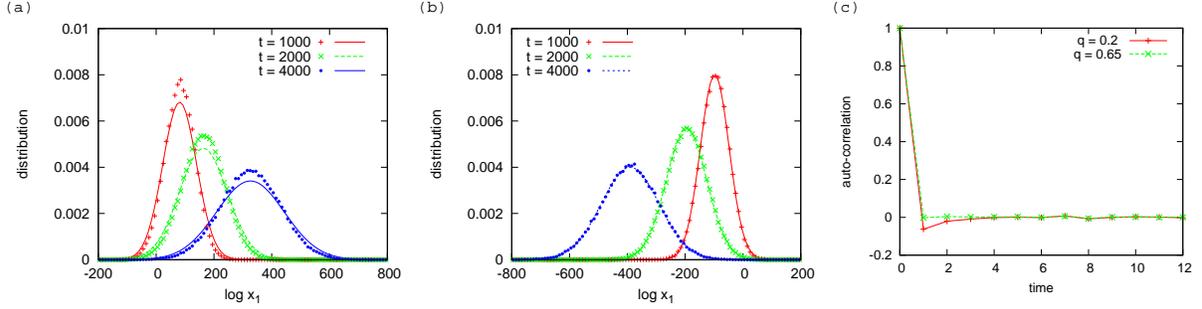}%
\caption{Evolution of probability distribution of $\ln x_1(t)$
for (a) $q=0.2$ and (b) $q=0.65$, 
when $t=1000, 2000, \mbox{ and } 4000$. 
The other parameters are $m_+=3$, $m_-=0.01$, $c=0$, $p=0.3$, and $s=0.5$.
The crosses stand for the distribution obtained by 100000 stochastic
realizations. 
The lines stand for the approximation with neglecting time correlation
of the effective growth rates.
(c) The temporal correlation functions of the effective growth rates
of $x_1(t)$ in the cases (a) and (b). 
We find a slight negative correlation at short time lags and rapid 
decay to zero. 
The negative correlation implies that the variance of the actual 
probability distribution tends to be smaller than 
the approximation with neglecting temporal correlation of the effective
growth rate. (Color online) \label{f2}}
\end{figure}

Because $x_1(t)$ and $x_2(t)$ obey the same stochastic process, $\ln x_2(T)$
follows the same normal distribution as $\ln x_1(T)$.
At a glance, the difference $\ln x_2(T)-\ln x_1(T)$ looks like to follow 
a normal distribution. 
But this insight is not correct. 
Recall that $\ln r(T)=\ln x_2(T)-\ln x_1(T)$. 
We have concluded that this distribution has a complicated form
(as is seen in Fig.~1). 
This indicates that the two variables  $x_1(T)$ and $x_2(T)$ are not
independent of each other but have a complex relationship. 
Thus, the simultaneous distribution of  $x_1(T)$ and $x_2(T)$
has a complicated self-similar structure. 
For the well-mixed case ($qs=1-q$), $x_1(t)$ and $x_2(t)$ simply coincide. 
However, unless the populations are well-mixed,  $\ln x_1(T)+\ln x_2(T)$
does not follow a normal distribution and furthermore resulting
$x_1(T)+x_2(T)$ does not follow a log-normal distribution. 
The current result holds regardless of the shape of the distribution of
$m_1(t)$ and $m_2(t)$, since it is not because 
of the restriction that $m_1(t)$ and $m_2(t)$ can 
have only two values. 

Which population survives is assessed by the ensemble average of the 
long-term population growth 
\begin{equation}
\left\langle \frac{1}{T}\ln x_1(T)
\right\rangle .
\label{e14}
\end{equation}
For the well-mixed case ($qs=1-q$), 
the long-term growth rate (\ref{e14}) is rewritten by the average of 
local growth rates
\begin{equation}
\begin{array}{lcl}
 \displaystyle
\left\langle \frac{1}{T}\ln x_1(T)\right\rangle
&=&p_{++}\ln(2m_+)+p_{--}\ln(2m_-)\\
& &+2p_{+-}\ln(m_++m_-)+\ln(1-q).
\end{array}
\label{e15}
\end{equation}
Contrarily, for the isolated case ($q=0$),
it is calculated as $(1-p)\ln m_+ + p \ln m_-$.
For general cases, however, the long-term growth rate (\ref{e14})
cannot be written in a simple form. 
To obtain the value of (\ref{e14}), we need to calculate (\ref{e13})
by the stationary distribution of (\ref{e9}) with the help of computer.

Figure 3 gives a comparison between numerical simulations and 
analytical results. 
If the long-term growth rate (\ref{e14}) for $q=0$ is negative,
the population cannot survive in a single habitat. 
As shown in Fig.~3(a), the growth rate increases with the 
migration rate $q$ until the optimal migration rate $q^*$,
where the long-term growth rate has the maximal value. 
Consequently, the dispersal is advantageous for the populations to persist.
Figure 3(a) shows that the effect of dispersal is strong (weak) 
when the environments of habitats have a negative (positive) correlation. 
In Fig.~3(b), we plot the optimal migration rate $q^*$ 
as a function of survival rate $s$ for five values of $p$. 
The curves for $p$ and $1-p$ coincide, 
because of the symmetry between the two habitats. 
Although the optimal migration rate $q^*$ decreases with decreasing $s$,
$q^*$ remains finite even for relatively small $s$. 
Note that the average $\langle \frac{1}{T}\ln x_1(T)\rangle$ is 
the logarithm of the geometric average of $x_1(T)$. Since $x_1(T)$ 
follows a lognormal distribution, the geometric average of $x_1(T)$
coincides with the median of $x_1(T)$. 
This means that when the long term population growth (eq.~(\ref{e14}))
is positive (negative), the population grows (decays) with a 
probability of more than half.
\begin{figure}
\includegraphics[width=\linewidth]{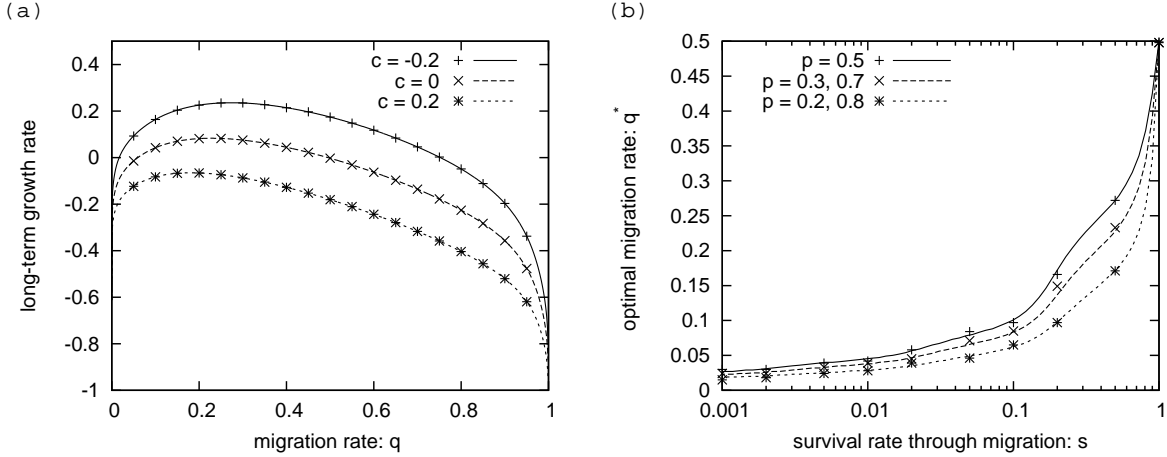}%
\caption{(a) Long term population growth rate is shown as a function of 
migration rate q for three values of correlation ($c=-0.2, 0, 0.2$). 
The other parameters are set as $m_+=3$, $m_-=0.01$, $p=0.3$, and $s=0.5$.
(b) The optimal migration rate is shown as a function of survival rate 
through migration s for $p=0.2,03,0.5,0.7,0.8$. 
The curve is obtained by (\ref{e13}) and the stationary distribution
$\rho_*(r)$. 
The crosses represent the numerical result of (\ref{e14}) 
for $T=10000$, averaged over 10000 stochastic realizations.
\label{f3}}
\end{figure}

Risk-spreading phenomena have been treated in biology and economics 
applying discrete stochastic linear metapolulation models. Most these 
studies are based on numerical approaches, except well-mixed cases that 
can be solved by simple algebra. Here we have solved analytically the 
cases of non-well-mixed populations. Our approaches can be widely 
applicable for forecasting long term trends not only in physical 
phenomena, but also in population biology and economic forecasting. 
Even complex stochastic problems may be tractable and solved analytically 
in this approach.

We analyzed the case of two habitats for simplicity. 
In the case of more than two habitats,
the qualitative prediction should be same as in the case 
two habitats.
Here we could expect the same 
The above results are expanded to the case of networks with more
than two habitats. 
Here, the ratios of populations among habitats 
have a complicated self-similar stationary distribution,
while each population always has a log-normal distribution. 
In other words, the simultaneous distribution has a complicated structure,
but its marginal distributions always follow a log-normal distribution. 
However, the numerical calculations become terribly cumbersome 
even in the case of three habitats. 
We should note here that this problem of multiple habitats may be 
treated in the framework of metapopulation dynamics
because some analytical solutions are already acquired\cite{cv1,cv2}.
We may also apply the current model to the studies of 
horizontal gene transfer in two distinctive environmental states\cite{no1,no2}.
We can also add a nonlinear effect to the current linear model.
Then, the stable distribution departs from a log-normal distribution.  
For example, the density dependent effect prevents a population from 
increasing to an infinitely large number. 
On the other hand, it is well-known that a finite injection leads to 
a power law distribution \cite{r17,r18,r19}.  
Although these cases have not been investigated here, 
we can conjecture that the simultaneous distribution may still show
have complicated self-similarity. 
In addition, some environmental factors are naturally correlated in 
successive times, resulting in the temporal correlations in local 
growth rates. These cases are treated numerically \cite{r7}, 
because temporal correlations are extremely complex and highly 
tedious to treat analytically. 
The current method may be applicable to solve these models analytically 
and we expect no qualitative differences from the current results.
We also find robustness in environmental parameters
$m_+$, $m_-$, $p$ and $c$. 
We generally could conclude that the stochasticity always 
promotes dispersal to ensure the long term survival 
even if the cost of migration is considerably high.

\begin{acknowledgments}
This work was supported by grants-in-aid from the Ministry of Education,
Culture, Sports, Science and Technology of Japan to S. M. (nos.24500273)
 and J. Y. (nos.22370010 and 22255004).
A part of numerical computation in this work was carried out at the 
Yukawa Institute Computer Facility.
\end{acknowledgments}

\bibliography{ref}

\providecommand{\noopsort}[1]{}\providecommand{\singleletter}[1]{#1}%
\begin{thebibliography}{23}%
\makeatletter
\providecommand \@ifxundefined [1]{%
 \@ifx{#1\undefined}
}%
\providecommand \@ifnum [1]{%
 \ifnum #1\expandafter \@firstoftwo
 \else \expandafter \@secondoftwo
 \fi
}%
\providecommand \@ifx [1]{%
 \ifx #1\expandafter \@firstoftwo
 \else \expandafter \@secondoftwo
 \fi
}%
\providecommand \natexlab [1]{#1}%
\providecommand \enquote  [1]{``#1''}%
\providecommand \bibnamefont  [1]{#1}%
\providecommand \bibfnamefont [1]{#1}%
\providecommand \citenamefont [1]{#1}%
\providecommand \href@noop [0]{\@secondoftwo}%
\providecommand \href [0]{\begingroup \@sanitize@url \@href}%
\providecommand \@href[1]{\@@startlink{#1}\@@href}%
\providecommand \@@href[1]{\endgroup#1\@@endlink}%
\providecommand \@sanitize@url [0]{\catcode `\\12\catcode `\$12\catcode
  `\&12\catcode `\#12\catcode `\^12\catcode `\_12\catcode `\%12\relax}%
\providecommand \@@startlink[1]{}%
\providecommand \@@endlink[0]{}%
\providecommand \url  [0]{\begingroup\@sanitize@url \@url }%
\providecommand \@url [1]{\endgroup\@href {#1}{\urlprefix }}%
\providecommand \urlprefix  [0]{URL }%
\providecommand \Eprint [0]{\href }%
\providecommand \doibase [0]{http://dx.doi.org/}%
\providecommand \selectlanguage [0]{\@gobble}%
\providecommand \bibinfo  [0]{\@secondoftwo}%
\providecommand \bibfield  [0]{\@secondoftwo}%
\providecommand \translation [1]{[#1]}%
\providecommand \BibitemOpen [0]{}%
\providecommand \bibitemStop [0]{}%
\providecommand \bibitemNoStop [0]{.\EOS\space}%
\providecommand \EOS [0]{\spacefactor3000\relax}%
\providecommand \BibitemShut  [1]{\csname bibitem#1\endcsname}%
\let\auto@bib@innerbib\@empty
\bibitem [{\citenamefont {Levin}(1976)}]{r1}%
  \BibitemOpen
  \bibfield  {author} {\bibinfo {author} {\bibfnamefont {S.}~\bibnamefont
  {Levin}},\ }\href@noop {} {\bibfield  {journal} {\bibinfo  {journal} {Annu.\
  Rev.\ Ecol.\ Evol.\ Syst.}\ }\textbf {\bibinfo {volume} {7}},\ \bibinfo
  {pages} {287} (\bibinfo {year} {1976})}\BibitemShut {NoStop}%
\bibitem [{\citenamefont {Hastings}(1983)}]{r2}%
  \BibitemOpen
  \bibfield  {author} {\bibinfo {author} {\bibfnamefont {A.}~\bibnamefont
  {Hastings}},\ }\href@noop {} {\bibfield  {journal} {\bibinfo  {journal}
  {Theor.\ Popul.\ Biol.}\ }\textbf {\bibinfo {volume} {24}},\ \bibinfo {pages}
  {244} (\bibinfo {year} {1983})}\BibitemShut {NoStop}%
\bibitem [{\citenamefont {Williams}\ and\ \citenamefont {Hastings}(2011)}]{r3}%
  \BibitemOpen
  \bibfield  {author} {\bibinfo {author} {\bibfnamefont {P.~D.}\ \bibnamefont
  {Williams}}\ and\ \bibinfo {author} {\bibfnamefont {A.}~\bibnamefont
  {Hastings}},\ }\href@noop {} {\bibfield  {journal} {\bibinfo  {journal}
  {Proc.\ R.\ Soc.\ B 278}\ }\textbf {\bibinfo {volume} {278}},\ \bibinfo
  {pages} {1281} (\bibinfo {year} {2011})}\BibitemShut {NoStop}%
\bibitem [{\citenamefont {Jansen}\ and\ \citenamefont {Yoshimura}(1998)}]{r4}%
  \BibitemOpen
  \bibfield  {author} {\bibinfo {author} {\bibfnamefont {V.~A.~A.}\
  \bibnamefont {Jansen}}\ and\ \bibinfo {author} {\bibfnamefont
  {J.}~\bibnamefont {Yoshimura}},\ }\href@noop {} {\bibfield  {journal}
  {\bibinfo  {journal} {Proc.\ Natl.\ Acad.\ Sci.\ USA}\ }\textbf {\bibinfo
  {volume} {95}},\ \bibinfo {pages} {3696} (\bibinfo {year}
  {1998})}\BibitemShut {NoStop}%
\bibitem [{\citenamefont {Roy}\ \emph {et~al.}(2005)\citenamefont {Roy},
  \citenamefont {Holt},\ and\ \citenamefont {Barfield}}]{r5}%
  \BibitemOpen
  \bibfield  {author} {\bibinfo {author} {\bibfnamefont {M.}~\bibnamefont
  {Roy}}, \bibinfo {author} {\bibfnamefont {R.~D.}\ \bibnamefont {Holt}}, \
  and\ \bibinfo {author} {\bibfnamefont {M.}~\bibnamefont {Barfield}},\
  }\href@noop {} {\bibfield  {journal} {\bibinfo  {journal} {Am.\ Nat.}\
  }\textbf {\bibinfo {volume} {166}},\ \bibinfo {pages} {246} (\bibinfo {year}
  {2005})}\BibitemShut {NoStop}%
\bibitem [{\citenamefont {Benaim}\ and\ \citenamefont {Schreiber}(2009)}]{r6}%
  \BibitemOpen
  \bibfield  {author} {\bibinfo {author} {\bibfnamefont {M.}~\bibnamefont
  {Benaim}}\ and\ \bibinfo {author} {\bibfnamefont {S.~J.}\ \bibnamefont
  {Schreiber}},\ }\href@noop {} {\bibfield  {journal} {\bibinfo  {journal}
  {Theor.\ Popul.\ Biol.}\ }\textbf {\bibinfo {volume} {76}},\ \bibinfo {pages}
  {19} (\bibinfo {year} {2009})}\BibitemShut {NoStop}%
\bibitem [{\citenamefont {Schreiber}(2010)}]{r7}%
  \BibitemOpen
  \bibfield  {author} {\bibinfo {author} {\bibfnamefont {S.~J.}\ \bibnamefont
  {Schreiber}},\ }\href@noop {} {\bibfield  {journal} {\bibinfo  {journal}
  {Proc.\ R.\ Soc.\ B}\ }\textbf {\bibinfo {volume} {277}},\ \bibinfo {pages}
  {1907} (\bibinfo {year} {2010})}\BibitemShut {NoStop}%
\bibitem [{\citenamefont {Matthews}\ and\ \citenamefont {Gonzalez}(2007)}]{r8}%
  \BibitemOpen
  \bibfield  {author} {\bibinfo {author} {\bibfnamefont {D.~P.}\ \bibnamefont
  {Matthews}}\ and\ \bibinfo {author} {\bibfnamefont {A.}~\bibnamefont
  {Gonzalez}},\ }\href@noop {} {\bibfield  {journal} {\bibinfo  {journal}
  {Ecology}\ }\textbf {\bibinfo {volume} {88}},\ \bibinfo {pages} {2848}
  (\bibinfo {year} {2007})}\BibitemShut {NoStop}%
\bibitem [{\citenamefont {Fernholz}\ and\ \citenamefont {B.Shay}(1982)}]{r9}%
  \BibitemOpen
  \bibfield  {author} {\bibinfo {author} {\bibfnamefont {R.}~\bibnamefont
  {Fernholz}}\ and\ \bibinfo {author} {\bibnamefont {B.Shay}},\ }\href@noop {}
  {\bibfield  {journal} {\bibinfo  {journal} {J.\ Finance}\ }\textbf {\bibinfo
  {volume} {37}},\ \bibinfo {pages} {615} (\bibinfo {year} {1982})}\BibitemShut
  {NoStop}%
\bibitem [{\citenamefont {Luenberger}(1998)}]{r10}%
  \BibitemOpen
  \bibfield  {author} {\bibinfo {author} {\bibfnamefont {D.~G.}\ \bibnamefont
  {Luenberger}},\ }\href@noop {} {\emph {\bibinfo {title} {Investment
  science}}}\ (\bibinfo  {publisher} {Prentice Hall},\ \bibinfo {year}
  {1998})\BibitemShut {NoStop}%
\bibitem [{\citenamefont {Dempster}(2007)}]{r11}%
  \BibitemOpen
  \bibfield  {author} {\bibinfo {author} {\bibfnamefont {M.~A.~H.}\
  \bibnamefont {Dempster}},\ }\href@noop {} {\bibfield  {journal} {\bibinfo
  {journal} {Quant. Finance}\ }\textbf {\bibinfo {volume} {7}},\ \bibinfo
  {pages} {151} (\bibinfo {year} {2007})}\BibitemShut {NoStop}%
\bibitem [{\citenamefont {Kelly}(1956)}]{r12}%
  \BibitemOpen
  \bibfield  {author} {\bibinfo {author} {\bibfnamefont {J.~L.}\ \bibnamefont
  {Kelly}},\ }\href@noop {} {\bibfield  {journal} {\bibinfo  {journal} {Bell
  Syst.\ Tech.\ J.}\ }\textbf {\bibinfo {volume} {35}},\ \bibinfo {pages} {917}
  (\bibinfo {year} {1956})}\BibitemShut {NoStop}%
\bibitem [{\citenamefont {Lasota}\ and\ \citenamefont {Mackey}(1985)}]{r15}%
  \BibitemOpen
  \bibfield  {author} {\bibinfo {author} {\bibfnamefont {A.}~\bibnamefont
  {Lasota}}\ and\ \bibinfo {author} {\bibfnamefont {M.~C.}\ \bibnamefont
  {Mackey}},\ }\href@noop {} {\emph {\bibinfo {title} {Probabilistic Properties
  of Deterministic Systems}}}\ (\bibinfo  {publisher} {Cambridge University
  Press},\ \bibinfo {year} {1985})\BibitemShut {NoStop}%
\bibitem [{\citenamefont {Morita}\ and\ \citenamefont {Chawanya}(2002)}]{r16}%
  \BibitemOpen
  \bibfield  {author} {\bibinfo {author} {\bibfnamefont {S.}~\bibnamefont
  {Morita}}\ and\ \bibinfo {author} {\bibfnamefont {T.}~\bibnamefont
  {Chawanya}},\ }\href@noop {} {\bibfield  {journal} {\bibinfo  {journal}
  {Phys.\ Rev.\ E}\ }\textbf {\bibinfo {volume} {65}},\ \bibinfo {pages}
  {046201} (\bibinfo {year} {2002})}\BibitemShut {NoStop}%
\bibitem [{\citenamefont {Metz}\ \emph {et~al.}(1983)\citenamefont {Metz},
  \citenamefont {de~Jong},\ and\ \citenamefont {Klinkhamer}}]{r13}%
  \BibitemOpen
  \bibfield  {author} {\bibinfo {author} {\bibfnamefont {J.~A.~J.}\
  \bibnamefont {Metz}}, \bibinfo {author} {\bibfnamefont {T.~J.}\ \bibnamefont
  {de~Jong}}, \ and\ \bibinfo {author} {\bibfnamefont {P.~G.~L.}\ \bibnamefont
  {Klinkhamer}},\ }\href@noop {} {\bibfield  {journal} {\bibinfo  {journal}
  {Oecologia}\ }\textbf {\bibinfo {volume} {57}},\ \bibinfo {pages} {166}
  (\bibinfo {year} {1983})}\BibitemShut {NoStop}%
\bibitem [{\citenamefont {Bascompte}\ \emph {et~al.}(2002)\citenamefont
  {Bascompte}, \citenamefont {Possingham},\ and\ \citenamefont
  {Roughgarden}}]{r14}%
  \BibitemOpen
  \bibfield  {author} {\bibinfo {author} {\bibfnamefont {J.}~\bibnamefont
  {Bascompte}}, \bibinfo {author} {\bibfnamefont {H.}~\bibnamefont
  {Possingham}}, \ and\ \bibinfo {author} {\bibfnamefont {J.}~\bibnamefont
  {Roughgarden}},\ }\href@noop {} {\bibfield  {journal} {\bibinfo  {journal}
  {Am.\ Nat.}\ }\textbf {\bibinfo {volume} {159}},\ \bibinfo {pages} {128}
  (\bibinfo {year} {2002})}\BibitemShut {NoStop}%
\bibitem [{\citenamefont {Colizza}\ and\ \citenamefont
  {Vespignani}(2007)}]{cv1}%
  \BibitemOpen
  \bibfield  {author} {\bibinfo {author} {\bibfnamefont {V.}~\bibnamefont
  {Colizza}}\ and\ \bibinfo {author} {\bibfnamefont {A.}~\bibnamefont
  {Vespignani}},\ }\href@noop {} {\bibfield  {journal} {\bibinfo  {journal}
  {Phys.\ Rev.\ Lett.}\ }\textbf {\bibinfo {volume} {99}},\ \bibinfo {pages}
  {148701} (\bibinfo {year} {2007})}\BibitemShut {NoStop}%
\bibitem [{\citenamefont {Colizza}\ and\ \citenamefont
  {Vespignani}(2008)}]{cv2}%
  \BibitemOpen
  \bibfield  {author} {\bibinfo {author} {\bibfnamefont {V.}~\bibnamefont
  {Colizza}}\ and\ \bibinfo {author} {\bibfnamefont {A.}~\bibnamefont
  {Vespignani}},\ }\href@noop {} {\bibfield  {journal} {\bibinfo  {journal}
  {J.\ Theor.\ Biol.}\ }\textbf {\bibinfo {volume} {251}},\ \bibinfo {pages}
  {450} (\bibinfo {year} {2008})}\BibitemShut {NoStop}%
\bibitem [{\citenamefont {Novozhilov}\ \emph {et~al.}(2005)\citenamefont
  {Novozhilov}, \citenamefont {Karev},\ and\ \citenamefont {Koonin}}]{no1}%
  \BibitemOpen
  \bibfield  {author} {\bibinfo {author} {\bibfnamefont {A.~S.}\ \bibnamefont
  {Novozhilov}}, \bibinfo {author} {\bibfnamefont {G.~P.}\ \bibnamefont
  {Karev}}, \ and\ \bibinfo {author} {\bibfnamefont {E.~V.}\ \bibnamefont
  {Koonin}},\ }\href@noop {} {\bibfield  {journal} {\bibinfo  {journal} {Mol.\
  Biol.\ Evol}\ }\textbf {\bibinfo {volume} {22}},\ \bibinfo {pages} {1721}
  (\bibinfo {year} {2005})}\BibitemShut {NoStop}%
\bibitem [{\citenamefont {Thattai}\ and\ \citenamefont {van
  Oudenaarden}(2004)}]{no2}%
  \BibitemOpen
  \bibfield  {author} {\bibinfo {author} {\bibfnamefont {M.}~\bibnamefont
  {Thattai}}\ and\ \bibinfo {author} {\bibfnamefont {A.}~\bibnamefont {van
  Oudenaarden}},\ }\href@noop {} {\bibfield  {journal} {\bibinfo  {journal}
  {Genetics}\ }\textbf {\bibinfo {volume} {167}},\ \bibinfo {pages} {523}
  (\bibinfo {year} {2004})}\BibitemShut {NoStop}%
\bibitem [{\citenamefont {Levy}\ and\ \citenamefont {Solomon}(1996)}]{r17}%
  \BibitemOpen
  \bibfield  {author} {\bibinfo {author} {\bibfnamefont {M.}~\bibnamefont
  {Levy}}\ and\ \bibinfo {author} {\bibfnamefont {S.}~\bibnamefont {Solomon}},\
  }\href@noop {} {\bibfield  {journal} {\bibinfo  {journal} {Int.\ J.\ Mod.\
  Phys.\ C}\ }\textbf {\bibinfo {volume} {7}},\ \bibinfo {pages} {595}
  (\bibinfo {year} {1996})}\BibitemShut {NoStop}%
\bibitem [{\citenamefont {Takayasu}\ \emph {et~al.}(1997)\citenamefont
  {Takayasu}, \citenamefont {Sato},\ and\ \citenamefont {Takayasu}}]{r18}%
  \BibitemOpen
  \bibfield  {author} {\bibinfo {author} {\bibfnamefont {H.}~\bibnamefont
  {Takayasu}}, \bibinfo {author} {\bibfnamefont {A.~H.}\ \bibnamefont {Sato}},
  \ and\ \bibinfo {author} {\bibfnamefont {M.}~\bibnamefont {Takayasu}},\
  }\href@noop {} {\bibfield  {journal} {\bibinfo  {journal} {Phys.\ Rev.\
  Lett.}\ }\textbf {\bibinfo {volume} {79}},\ \bibinfo {pages} {966} (\bibinfo
  {year} {1997})}\BibitemShut {NoStop}%
\bibitem [{\citenamefont {Sornette}(1998)}]{r19}%
  \BibitemOpen
  \bibfield  {author} {\bibinfo {author} {\bibfnamefont {D.}~\bibnamefont
  {Sornette}},\ }\href@noop {} {\bibfield  {journal} {\bibinfo  {journal}
  {Phys.\ Rev.\ E}\ }\textbf {\bibinfo {volume} {57}},\ \bibinfo {pages} {4811}
  (\bibinfo {year} {1998})}\BibitemShut {NoStop}%
\end{thebibliography}%

\end{document}